%% file: main_new.tex
\documentclass{nature}

\usepackage{import}
\usepackage{amsmath}
\usepackage{graphicx}
\usepackage{xfrac}
\usepackage{multirow}
\usepackage{subfigure}
\usepackage{mathtools}
\usepackage[colorlinks=true,linkcolor=blue,citecolor=black,urlcolor=blue]{hyperref}
\usepackage{soul,xcolor}
\usepackage{authblk}
\usepackage[labelfont=bf]{caption}
\usepackage[figurename=Figure ]{caption}
\usepackage{siunitx}
\usepackage{scalefnt}

\usepackage{lineno}

\makeatletter

\let\saved@includegraphics\includegraphics
\AtBeginDocument{\let\includegraphics\saved@includegraphics}
\renewenvironment*{figure}{\@float{figure}}{\end@float}
\makeatother

\usepackage{color}
\usepackage{dsfont}
\usepackage[normalem]{ulem}
\usepackage{dcolumn}
\usepackage{blindtext}
\usepackage{outlines}
\usepackage{enumitem}
\usepackage{soul}
\setlist[enumerate,2]{label=\roman*)}
\setlist[enumerate,3]{label=\alph*)}
\usepackage{multirow}

\newcommand{\iu}{\mathrm{i}}

\newcolumntype{d}[1]{D{.}{.}{#1}}
\definecolor{orange}{rgb}{1,0.5,0}
\definecolor{darkgreen}{RGB}{0,100,0}

\setstcolor{red}

\begin{document}

\title{A new view on the origin of zero-bias anomalies of Co atoms atop noble metal surfaces}

\author{Juba Bouaziz$^\ast$, Filipe Souza Mendes Guimar\~aes, Samir Lounis$^\ast$}
\affil{Peter Gr\"unberg Institut and Institute for Advanced Simulation, 
Forschungszentrum  J\"ulich and JARA, 52425 J\"ulich, Germany}

\affil{$^\ast$  E-mails: j.bouaziz@fz-juelich.de,  s.lounis@fz-juelich.de.}

\date{}

\maketitle


\begin{abstract}
Many-body phenomena are paramount in physics. 
In condensed matter, their hallmark is considerable on a wide range of material characteristics spanning electronic, magnetic, thermodynamic and transport properties. 
They potentially imprint non-trivial signatures in spectroscopic measurements, such as those assigned to Kondo, excitonic and polaronic features, whose emergence depends on the involved degrees of freedom. 
Here, we address systematically zero-bias anomalies detected by scanning tunneling spectroscopy on Co atoms deposited on Cu, Ag and Au(111) substrates, which remarkably are almost identical to those obtained from first-principles. 
These features originate from gaped spin-excitations induced by a finite magnetic anisotropy energy, in contrast to the usual widespread interpretation relating them to Kondo resonances. 
Resting on relativistic time-dependent density functional and many-body perturbation theories, we furthermore unveil a new many-body feature, the spinaron, resulting from the interaction of electrons and spin-excitations localizing electronic states in a well defined energy.

\end{abstract}

\section*{Introduction}

\begin{figure}
	\centering
	\includegraphics[width=\textwidth]{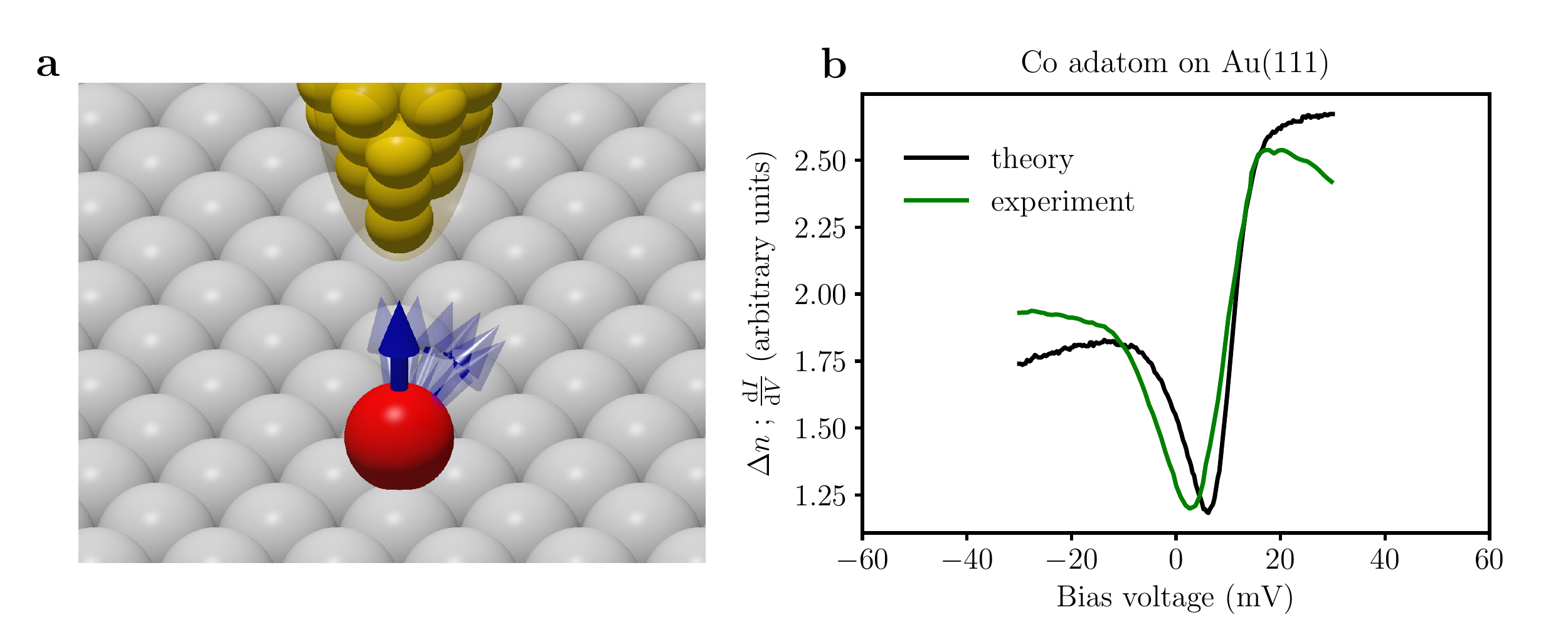}
	\caption{\textbf{Scanning tunneling spectroscopy probing the differential conductance of a Co adatom.} 
	(\textbf{a}) Illustration of the tip of a microscope scanning the surface of Au(111) on which an fcc Co adatom is deposited. 
	(\textbf{b}) The differential conductance, $dI/dV$, measured at T = \SI{6}{\kelvin}~\cite{Schneider:2005} compared to  first-principles results. 
	As deduced from the change of the LDOS owing to the presence of spin-excitations, $\Delta n$, the zero-bias anomaly stems from gaped spin-excitations and the presence of a many-body bound state, a spinaron,  at positive bias voltage. The experimental data  adapted with permission from IOP Publishing -- Japan Society of Applied Physics -- Copyright (2005).
	} \label{panel1}
\end{figure}

Signatures of many-body phenomena in solid state physics are diverse~\cite{keimer:2017,byrnes:2014,Pommier:2019,Pohlit:2018,Madhavan:1998}. One of them is the Kondo effect emerging from the 
interaction between the sea of electrons in a metal and the magnetic moment of an atom~\cite{Kondo:1964,hewson_1993}, whose signature is expected below a characteristic Kondo temperature $T_K$. 
One of its manifestations is a resistivity minimum followed by a strong increase upon reducing the temperature, as initially observed in metals doped with a low concentration of magnetic impurities~\cite{Monod_res:1967}. 
When the latter are deposited on surfaces, they can develop Kondo resonances evinced by zero-bias anomalies, with various Fano-shapes~\cite{Fano:1961,Ujsaghy:2000} that are detectable by scanning tunneling spectroscopy (STS), as shown schematically in Figure~\ref{panel1}a. The discovery of such low-energy spectroscopic features by pioneering STS measurements~\cite{Madhavan:1998,Knorr:2002ig,Heinrich:2004,Ternes:2008} opened an active research field striving to address and learn about many-body physics at the sub-nanoscale.
A seminal example is Co adatoms deposited on Cu, Ag and Au(111) surfaces~\cite{Madhavan:1998,Jamneala:2000,Madhavan:2001,Knorr:2002ig,Wahl:2004jy,Schneider:2005,Ternes:2008,Wahl:2009hs}, which  develop a dip in the transport spectra, with a minimum located at a positive bias voltage surrounded by steps from either sides (Figure~\ref{panel1}b). 
Although being commonly called Kondo resonances, the hallmarks of the Kondo effect have so far not been established for those particular Co atoms, i.e. the disappearance of the Kondo resonance at temperatures above $T_K$ and the splitting of the feature after applying a magnetic field~\cite{hewson_1993,Costi:2000,Nagaoka:2002,Heinrich:2004,Otte:2008}.  
A huge progress was made in advanced simulations combining quantum impurity solvers or even GW with density functional theory (DFT) addressing Kondo phenomena for various impurities (see e.g. Refs.~\citenum{Thygesen:2008,huang:2008,surer:2012,Jacob_2015,dang:2016}), often neglecting spin-orbit interaction. The electronic structure spectra of  realistic systems do not reproduce, in general, the experimental ones.

In the current work, we provide an alternative interpretation for the observed zero-bias anomalies in Co adatoms deposited on Cu, Ag and Au(111) surfaces, utilizing a recently developed framework resting on relativistic time-dependent DFT (TD-DFT) in conjunction with many-body perturbation theory (MBPT). Similar results were found for Co adatoms on Cu, Ag(001) surfaces and Ti adatom on Ag(001), which are shown in Supplementary Figure~1 for the sake of brevity. Our first-principles simulations indicate that the observed features find their origin in inelastic spin-excitations (SE), as known for other systems~\cite{Heinrich:2004,Hirjibehedin:2006,Rossier:2009,Samir:2010,Khajetoorians:2011,Chilian:2011,Khajetoorians:2013,Bryant:2013,Oberg:2014,Ternes:2015}, which are gaped SE owing to the magnetic anisotropy energy that favors the out-of-plane orientation of the Co moment. Therefore, the physics is dictated by relativistic effects introduced by the spin-orbit interaction. As illustrated in Figure~\ref{panel1}b, the resulting theoretical transport spectra are nearly identical to the experimental ones, advocating for a non-Kondo origin of the features.
This effect induces two steps, asymmetric in their height, originating from intrinsic spin-excitations, and leads to the typically observed shape in the differential conductance, thanks to the emergence in one side of the bias voltage of a new type of many-body feature: a bound state that we name spinaron, emanating from the interaction of the spin-excitation and electrons.  Finally, we propose possible experiments that enable the verification of the origin of the investigated zero-bias anomalies.

\section*{Results}

We compare our theoretical data to measurements obtained with low-temperature STS and proceed with a three-pronged approach for the first-principles simulations. 
We start from regular DFT calculations based on the full-electron Korringa-Kohn-Rostoker (KKR) Green function~\cite{Papanikolaou:2002,Bauer:2014} method, which is ideal to treat Co adatoms on metallic substrates. 
We continue by building the tensor of relativistic dynamical magnetic susceptibilities for the adatom-substrate complex, $\underline{\chi}(\omega)$, encoding the spectrum of SEs~\cite{Samir:2010,Samir:2015,Manuel:2015}. 
Finally, the many-body self-energy, $\underline{\Sigma}(\varepsilon)$, is computed accounting for the SE-electron interaction including the spin-orbit coupling. 
The Tersoff-Hammann approach~\cite{Tersoff1983} allows the access to the differential conductance via the ground-state LDOS decaying from the substrate to the vacuum, where the STS-tip is located, here assumed to be located at \SI{6.3}{\angstrom} above the adatom for the Cu(111) surface and \SI{7.1}{\angstrom} for the Ag and Au(111) ones.
This is then used to evaluate the renormalization of the differential conductance because of the SEs.
More details are given in the Methods section and Supplementary Note 1.

\begin{figure}
	\centering
	\includegraphics[width=\textwidth]{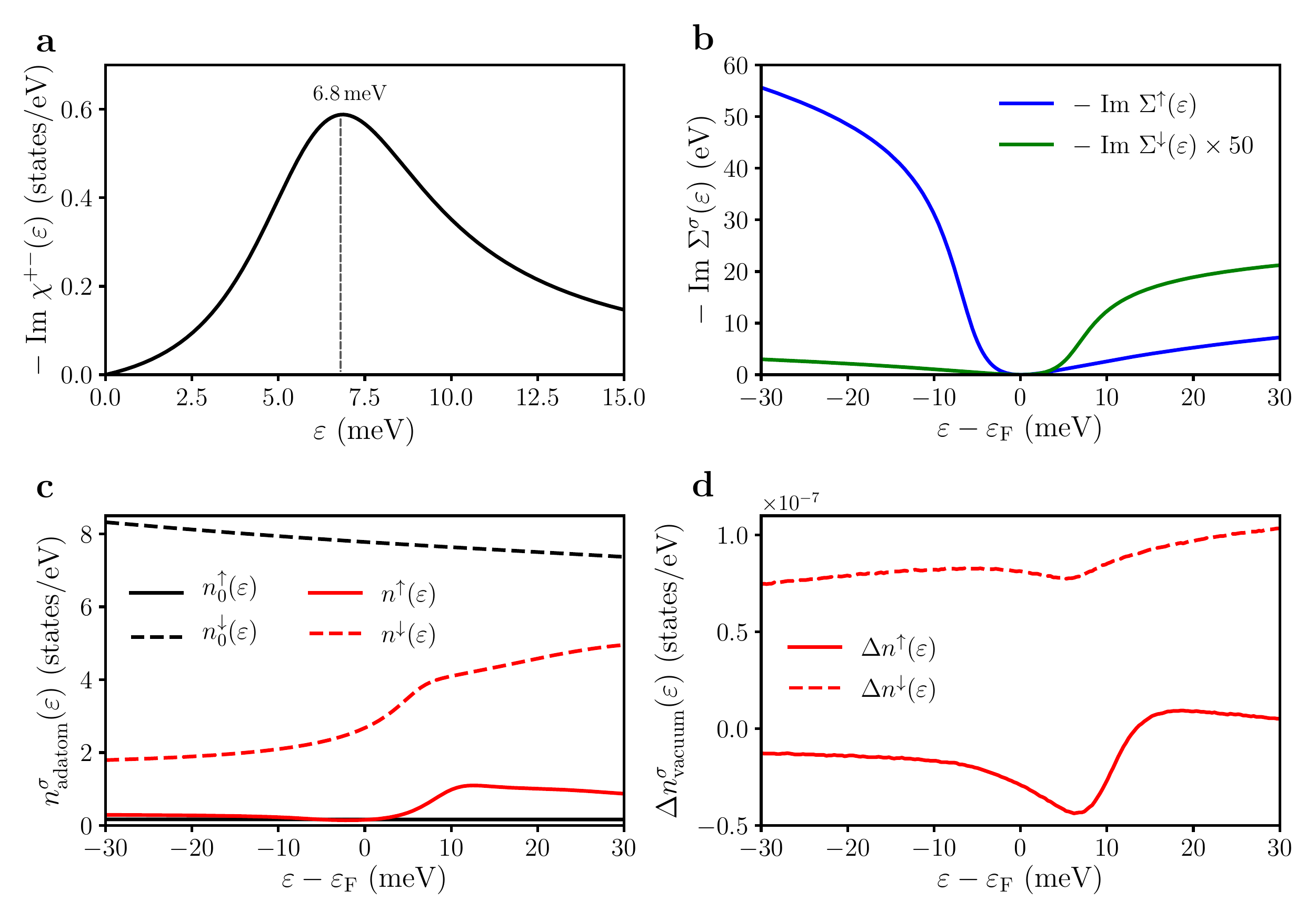}
	\caption{\textbf{Spin-excitations, self-energies and local density of states for Co adatom/Au(111).} 
	(\textbf{a}) Density of spin-excitations showing a broad resonance located at \SI{6.8}{\milli\electronvolt}, due to the adatom's magnetic anisotropy energy. 
	The lifetime of the spin-excitation is \SI{0.29}{\pico\second}, which is mainly settled by electron-hole excitations. 
	(\textbf{b}) Spin-resolved electronic self-energy $\Sigma$, which inherits the information on the presence of the spin-excitation by hosting asymmetric steps above and below the Fermi energy. $\uparrow$ ($\downarrow$) stands for majority-spin (minority-spin) . 
	(\textbf{c}) LDOS at the adatom site before ($n_0$) and after ($n$) accounting for the interaction between electrons and the spin-excitation.
	(\textbf{d}) Spin-resolved change in the LDOS ($\Delta n$) calculated in the vacuum above the adatom. The minority-spin channel shows a low-intensity feature above the Fermi energy as expected from the corresponding self-energy. 
	The large feature present in the majority-spin channel is composed of two steps, one originating from the intrinsic spin-excitation while the other is the spinaron, arising from the interaction of electrons and spin-excitations.
	} \label{panel2}
\end{figure}

\subsection{Zero-bias anomaly of Co adatom on Au(111).}
We discuss here the different ingredients leading to the spectrum shown in Figure~\ref{panel1}b, which was found to be in a remarkable agreement with the data of Ref.~\citenum{Schneider:2005}, in particular. The adatom on Au(111) surface carries a spin moment of $2.22 \,\mu_\text{B}$ and a relatively large orbital moment of $0.43\,\mu_\text{B}$. The easy axis of the Co magnetic moment is out-of-plane favored by a substantial magnetic anisotropy energy (MAE) of \SI{4.46}{\milli\electronvolt} (Table~\ref{gs_parameters}). This opens a gap in the SE spectrum, as illustrated in Figure~\ref{panel2}a, which shows the density of transversal SEs describing spin-flip processes, $-\frac{1}{\pi}\,\text{Im}\, \chi^{+-} (\omega)$. The SE arises at \SI{6.8}{\milli\electronvolt}, which is shifted from the expected ideal location, \SI{8}{\milli\electronvolt}, as obtained from $4\frac{\text{MAE}}{M_\text{spin}}$ because of dynamical corrections\cite{Manuel:2015}. As a result of electron-hole excitations of opposite spins~\cite{Lounis:2010,Lounis:2011}, the   lifetime $\tau$  of the SE is reduced down to \SI{0.29}{\pico\second} ($\tau = \frac{\hbar}{\Gamma}$, $\Gamma$ being the resonance width at half maximum). 
A simplified theory indicates that this effective damping is enhanced by the finite  LDOS at the Fermi energy, which settles the density of electron-hole excitations~\cite{Samir:2015}. 
The interaction of electrons and spin-excitations is incorporated in the so-called self-energy. 
It is represented by a complex quantity, with the real part shifting the energy of the electrons, and the imaginary part describing their inverse lifetimes.
The significant components of the self-energy are spin diagonal, considering that the contribution of the off-diagonal elements is negligible for the investigated C$_\text{3v}$-symmetric adatom-substrate systems (see Supplementary Note 1). 
These quantities are computed from the dynamical susceptibilities $\underline{\chi}(\omega)$ and the ground-state density $n_0(\varepsilon)$.
For instance, the imaginary part for a given spin channel ${\sigma}$,  $\text{Im}\,\underline{\Sigma}^{{\sigma}{\sigma}}(\varepsilon_\text{F}+V)$, is proportional to $\int_{0}^{-V} d\omega\,n_0^{\bar{\sigma}}(\varepsilon_\text{F}+V+\omega)\,\text{Im}\,\chi^{\sigma\bar{\sigma},\bar{\sigma}\sigma}(\omega)$, i.e. it is a convolution of the ground-state density, $n_0(\varepsilon)$, of the opposite spin-character and the SE density integrated over the bias voltage of interest. 
This quantity is plotted in Figure~\ref{panel2}b.
Two steps are present, one for each spin channel, located at positive (negative) bias voltage for the minority (majority) self-energy.
This is expected from the integration of the SE density over a resonance.
As the self-energy of a given spin-channel is proportional to the LDOS of the opposite spin-character, the majority-spin self-energy has a higher intensity than the minority one, as expected from adatom LDOS illustrated in Figure~\ref{panel2}c, which decreases substantially the lifetime of the majority-spin electrons compared to that of minority-spin type. 
\begin{table}
	\begin{center}
		\begin{tabular}{lcccc}
			\hline 
			Surface  & $\tau$ (\SI{}{\pico\second}) &MAE (\SI{}{\milli\electronvolt}) & $M_\text{spin}$($\mu_\text{B}$) & $M_\text{orb}$($\mu_\text{B}$) \\
		    \hline 
			Cu(111) & 0.098 & 4.29 & 2.02 & 0.47 \\
			\hline 
			Ag(111) & 0.088 & 3.27 & 2.21 & 0.70 \\
			\hline
			Au(111) & 0.073 & 4.46 & 2.22 & 0.44\\
			\hline
		\end{tabular}
		\caption{\textbf{Basic calculated properties of Co adatom on Cu, Ag and Au(111) surfaces.} Lifetime of the transverse spin excitations $\tau$, magneto-crystalline anisotropy (MAE), amplitude of the spin $M_\text{spin}$($\mu_\text{B}$) and orbital moments $M_\text{orb}$($\mu_\text{B}$). The positive sign of the MAE indicates that the magnetic moment points perpendicular to the three investigated substrates.} 
        \label{gs_parameters}
	\end{center}
\end{table}

The ground-state LDOS, $n^{\sigma}_{0}(\varepsilon)$, of Co adatom (depicted in Figure~\ref{panel2}c as black lines) varies very weakly for a bias voltage range of $\sim$ \SI{60}{\milli\electronvolt} at the vicinity of the Fermi energy. 
The LDOS is then renormalized by the SE-electron interaction upon solving the Dyson equation  
\begin{equation}
\underline{G}_\text{R}(\varepsilon) = 
[1 - \underline{G}(\varepsilon)\,\underline{\Sigma}(\varepsilon)]^{-1}\underline{G}(\varepsilon),
\label{Dyson-eq}
\end{equation} 
from which the renormalized LDOS is obtained by tracing over site, spin and angular momenta of the Green function: $n(\varepsilon) = -\frac{1}{\pi} \,\text{Im}\,\text{Tr}\,\underline{G}_\text{R}(\varepsilon)$.

At the adatom site (Figure~\ref{panel2}c), step-like features arise in the LDOS at the SE energy. The minority-spin LDOS hosts one single feature above the Fermi energy as expected from the corresponding self-energy. 
In contrast, the majority-spin LDOS is marked with an additional feature at positive voltage, which we identify as a many-body bound state --- a spinaron. One can recognize it (spinaron) either from a one-to-one comparison between the spin-resolved LDOS and the self-energy, as being a feature not present in the latter one (see Supplementary Figures~3 and 4) or from tracking the intersections of Green functions and self-energies leading to the vanishing of the denominator of Eq.~\eqref{Dyson-eq}. The presence of spin-fluctuations affect the electronic behavior in terms of the electron-SE interaction encoded in the self-energy. This additional interaction can act as an attractive potential permitting the localization of electrons in a finite energy window, giving rise to a bound state. The spinaron emerges then when the denominator of the Dyson equation, Eq.~\eqref{Dyson-eq}, cancels out, i.e. when $\text{Re}(\underline{G}\underline{\Sigma}) = 1$, which occurs for the $d_{z^2}$ orbital having the ideal symmetry to be detected by STS, as illustrated in Supplementary Figures~5 and 6. 
The spinaron is characterized by an energy and a lifetime (settled by $\text{Im}(\underline{G}\underline{\Sigma})$), both affected by the spin-orbit interaction, since it dictates the magnitude of the SE-gap defining the self-energy, and the electron-hole excitations.

The adatom electronic features decay into the vacuum, which are probed by the STS-tip in terms of the  differential conductance. The signature of the SE is better seen in the change of the vacuum LDOS, $\Delta n = n - n_0$, illustrated in Figure~\ref{panel1}b and being  spin-decomposed in Figure~\ref{panel2}d. 
One sees that the origin of the two steps and their asymmetry observed experimentally and theoretically is the concomitant contribution of the spin-excitation features and the spinaron. The signal is mainly emanating from the majority-spin LDOS, with the spinaron showing up as a step being higher than the one corresponding to the intrinsic SE below the Fermi energy. We note that the spinaron bears similarities to the spin polaron suggested to exist in halfmetallic ferromagnets~\cite{Irkhin:2007}. 
\begin{figure}
	\centering
	\includegraphics[width=\textwidth]{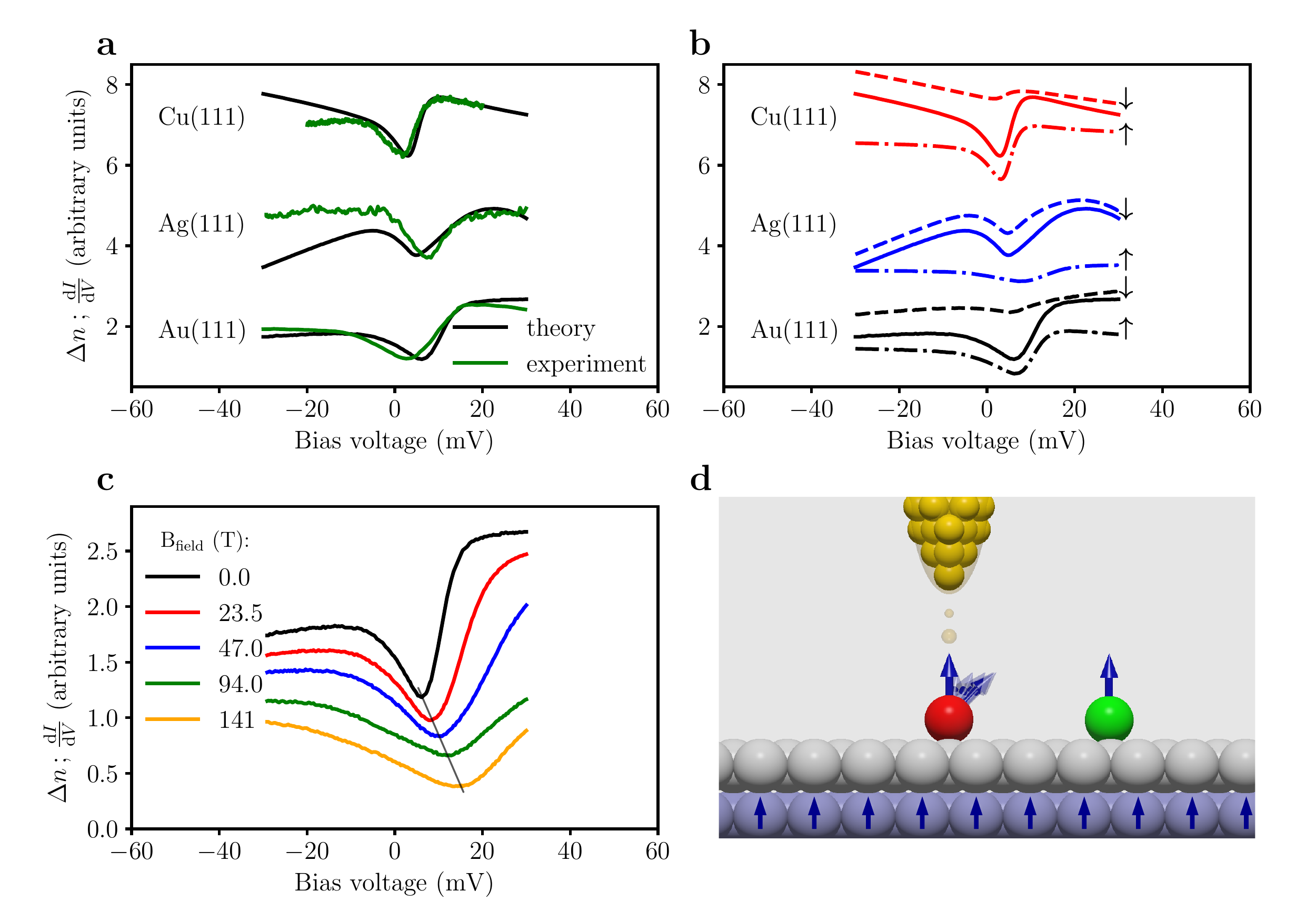}
	\caption{\textbf{Systematic tunneling transport spectra of Co adatoms on Cu, Ag and Au(111) surfaces.} 
	(\textbf{a}) Excellent agreement between the first-principles spectra and  those measured with STS at temperatures of \SI{1.2}{\kelvin}~\cite{Ternes:2008}, \SI{1.1}{\kelvin}~\cite{MoroLagares:2019ff}, \SI{6}{\kelvin}~\cite{Schneider:2005} 
	(\textbf{b}) Spin-resolved spectra indicating that spin-polarized spectroscopy can reshape the measured differential conductance, which would help to disentangle the various contributions to the spectra. The total (solid) signal is obtained by the sum of both spin channels, which were shifted for a better visualization and comparison.
	(\textbf{c}) Response of the zero-bias anomalies to magnetic fields. The spin-excitation gap opens while the dip moves to larger energies. 
	Large magnetic field are possible in some STS setups (\SI{14}{\tesla}~\cite{assig:2013} or even \SI{38}{\tesla}~\cite{tao:2017,rossi:2018}). 
	(\textbf{d}) Proximity effects can be used via neighboring adatoms and  by depositing thin films of noble metals on a ferromagnetic substrate, where the magnetic exchange interaction felt by the adatom acts as an effective magnetic field.  The experimental data in (\textbf{a}) adapted with permission from IOP Publishing for Cu~\cite{Ternes:2008} and from the Japan Society of Applied Physics -- Copyright (2005) for  Au~\cite{Schneider:2005} as well as from Nature Publishing Group for Ag~\cite{MoroLagares:2019ff}. 
	} \label{panel3}
\end{figure}

\subsection{Systematic study of Co adatoms on Cu, Ag, Au(111) surfaces.}
We performed a systematic comparison between simulated and experimental data and evidenced that the the spin-excitations combined with the spinaron are generic features for Co adatoms deposited on  Cu(111), Ag(111) and Au(111). 
The agreement shown in Figure~\ref{panel3}a is staggering, certifying that the result obtained for Au(111) surface is not accidental. Similarly to Au, the spinaron originates from the $d_{z^2}$ on Ag and Cu, conferring the right symmetry to be detected efficiently with STS (more details are provided in Supplementary Figures~5 and 6). 
This enforces our view that the experimentally observed zero-bias features for Co adatoms are captured by  gaped SEs. The energies and lifetimes $(\varepsilon_\text{spinaron},\tau_\text{spinaron})$ of the spinarons as obtained from the theoretical spectra in vacuum are: $( \SI{4.42}{\milli\electronvolt}, \SI{0.34}{\pico\second})$, $( \SI{12.6}{\milli\electronvolt}, \SI{0.20}{\pico\second})$ and $( \SI{9.41}{\milli\electronvolt}, \SI{0.20}{\pico\second})$ for Cu, Ag and Au(111), respectively, which are of the same order of magnitude than those of the intrinsic spin-excitations listed in Table~\ref{gs_parameters}.  Interestingly, the lifetimes of the latter excitations increase slightly on Cu and Ag surfaces when compared to that obtained on Au. Interestingly, the spectra obtained for Co/Cu(111) are in line with those reported in Ref.~\cite{Schweflinghaus:2014} based on a simplified theoretical approach (more details are provided in Supplementary Note 2).

Merino and Gunnarsson~\cite{Merino:2004} suggested that the surface states of the investigated substrates give rise to the particular shape of the low energy excitations. In the case of Ag(111), STS experiments showed the possible alteration of the tunneling signal depending on the scattering of the Ag surface state at surrounding defects and step edges~\cite{Moro-Lagares:2018}. When compared to other surfaces (see Supplementary Figure 1), our theory indicates that the surface states are important to enhance the overall signal in the vacuum while the main origin of the spectral anomalies of the isolated adatoms is a combination of the intrinsic spin-excitations signatures and the spinaron. The weight and shape of each of the features depend on the substrate and interference effects induced by decay of the electronic states of both the adatom and surface. Moreover Eq.~\eqref{Dyson-eq}, shows that both the step-like and peak-like features of the respective imaginary and real parts of the self-energy are mixed up, contributing to the signature of the observed low-energy anomalies (see Supplementary Figure~3). One sees in Figure~\ref{panel3}b, that the main difference between the three surfaces originates from intrinsic spin-excitation occurring in the minority-spin channel (positive bias voltage), which is for Ag more enhanced than the signal coming from the majority-spin channel carrying both the intrinsic spin-excitation (negative bias voltage) and the spinaron (positive bias voltage). When deposited on Cu and Au, the asymmetry between majority- and minority-spin channels switches. This is induced by the electronic structure of the adatom on the three surfaces (see Supplementary Figure~2). 
The LDOS at the Fermi energy of Co on Ag hosts a larger minority-spin DOS than on Cu and Au. 
The reason is the weaker hybridization strength between the electronic states of Co and the substrate, when compared to Cu or Au(111), which reduces the broadening of the minority-spin resonance on the former. 
The positions of the steps pertaining to the intrinsic SEs correlate with the magnitude of the MAE, which  favors the out-of-plane orientation of the Co magnetic moment on the three substrates as listed in Table~\ref{gs_parameters}. 

For the quantitative validation of the agreement between the theoretical and experimental data, we fit our data with the commonly used Fano-resonance formula for the differential conductance of Kondo resonances~\cite{Fano:1961,Ujsaghy:2000,Madhavan:2001}:
\begin{equation}
n(\varepsilon) = \mathcal{A}\,\frac{\left(\varepsilon+q\right)^2}{\varepsilon^2 + 1}\quad, 
\label{Fano_resonance_fit}
\end{equation}
with $\mathcal{A}$ being the amplitude of the signal and $q$ the coupling parameter. The latter plays an important role in the Fano formalism as it determines the shape and asymmetry of the STS-signal. 
$\varepsilon = (eV- E_{0})/k_\text{B}T^\text{eff}_\text{K}$ encodes the information regarding the effective Kondo temperature $T^\text{eff}_\text{K}$, as well as the bias voltage $V$ (with $k_\text{B}$ being the Boltzmann constant and $E_{0}$ the position of the investigated resonance). 
The fitted Fano-parameters are listed in Table~\ref{Kondo_parameters}. 
Astonishingly, the recovered effective temperatures are in perfect agreement with the ones obtained from experimental data.  

\begin{table}
	\begin{center}
		\begin{tabular}{lccccc}
			\hline 
			Surface     & $E_0$(\SI{}{\milli\electronvolt})&  $T^\text{eff}_\text{K}$(\SI{}{\kelvin}) & $T^\text{exp}_\text{K}$(\SI{}{\kelvin})  &  $q$ & $q^\text{exp}$ \\
		    \hline 
			Cu(111) & 3.74 & 37.3 & 44.9 [57~\cite{Schneider:2005}] &  0.42 & 0.38 [0.5~\cite{Ternes:2008}]\\
			\hline 
			Ag(111) & 4.71 & 89.4 & 73 [56~\cite{Moro-Lagares:2018,MoroLagares:2019ff}] & -0.05 & -0.004 [0.02$\pm$0.02~\cite{Moro-Lagares:2018,MoroLagares:2019ff}] \\
			\hline
			Au(111) & 10.61 & 67.5 & 91 [76$\pm$8~\cite{Schneider:2005}]  &  0.55 & 0.45 [0.7~\cite{Schneider:2005}]\\
		
			\hline
		\end{tabular}
		\caption{\textbf{Fitted Fano parameters for a single Co adatom deposited on Cu, Ag and Au(111) surfaces.} 
		The resonance formula used for fitting is given in Eq.~\eqref{Fano_resonance_fit}. 
		The effective Kondo temperature $T^\text{eff}_\text{K}$ and the coupling parameter $q$ extracted from our own fits of the experimental and theoretical spectra are compared to those published in Refs.~\citenum{Schneider:2005,Ternes:2008,Moro-Lagares:2018,MoroLagares:2019ff} (values between brackets).
		} 
        \label{Kondo_parameters}
	\end{center}
\end{table}

\subsection{Experimental proposals: Impact of spin-polarized tip, magnetic field and proximity-effects.}
As mentioned before, the Kondo origin of the low-energy spectral features has so far not been evidenced for the systems investigated in the current work. This is usually realized by performing temperature-dependent measurements and/or upon the application of a magnetic field, which would respectively result in a broadening of the anomalies and/or their splitting. However, the large broadening of the dip-like structure require improved energy resolutions than currently available, preventing the realization of such experimental studies. Here we address possible experiments that can further verify our predictions. 

Kondo resonances should not change when probed by a spin-polarized tip. Our spectra are however spin-dependent and thus we expect the alteration of their shape depending on the spin-polarization of the tip, $P_\text{tip} = \frac{n^\uparrow_\text{tip} - n^\downarrow_\text{tip}}{n^\uparrow_\text{tip} + n^\downarrow_\text{tip}}$, since the differential conductance is approximately proportional to $(1 + P_\text{tip}) n^\uparrow_\text{adatom} + (1 - P_\text{tip}) n^\downarrow_\text{adatom}$~\cite{Tersoff1983,Wortmann:2001}. 
Ultimately, manipulating the spin-polarization of the tip (see e.g. Ref.~\citenum{loth:2010}) would help spin-resolving the LDOS as depicted in Figure~\ref{panel3}b for the three investigated substrates.

Furthermore, the zero-bias dip is expected to split into two features for a traditional Kondo resonance once an external magnetic field is applied~\cite{hewson_1993}. 
Figure~\ref{panel3}c shows a completely different behavior. 
The field applied along the easy axis of the Co atoms yields an increase of the excitation gap, as expected, and of the spinaron energy  
(see the spin-resolved spectra in Fig. S1).
The interplay of the various features gives the impression that the observed dip drifts to energetically higher unoccupied states, which occurs because of  the presence of the spinaron. We note that applying a magnetic field in the direction perpendicular to the magnetic moment, would affect the excitation gap in a non-trivial way~\cite{Khajetoorians:2013}. 
A field of 14 Tesla is available in some STS setups~\cite{assig:2013} and can even reach 38 Tesla~\cite{tao:2017,rossi:2018}. 
Larger fields can be accessed effectively via magnetic-exchange-mediated proximity effect by either (i) bringing another magnetic atom to the vicinity of the probed adatom or (ii) depositing the probed adatom on a magnetic surface with a non-magnetic spacer in-between (see Figure~\ref{panel3}d). If the adjacent atom is non-magnetic, it can modify the MAE, which dictates the magnitude of the SE gap. If the MAE is reduced, the lifetime of the spin-excitations is expected to increase, since the amount of electron-hole excitations available in the respective energy range would decrease. This can then favor the monitoring of the impact of temperature and magnetic field on the zero-bias anomalies, helping to distinguish a Kondo behavior from the one emerging from spin-excitations. 

\section*{Discussion}

The zero-bias anomalies probed by low-temperature scanning tunneling spectroscopy on Co atoms deposited on Cu, Ag and Au (111) surfaces, usually identified as  Kondo resonances, are shown to be the hallmarks of gaped spin-excitations enhanced by the presence of spinarons. We note that there are other  examples of materials, such as quantum wires, where zero-bias anomalies have been challenged to be Kondo features~\cite{Chen:2009,Sarkozy:2009}. However, the adatoms investigated in the current work represent the most traditional systems, where the surface science community converges to the Kondo-related interpretation.
The gap of the spin-excitations is induced by the magnetic anisotropy energy of the Co adatom, defining the \SI{}{\milli\electronvolt} energy scale requested to excite the magnetic moment, and therefore its magnitude can be extracted from the position of the observed steps. Considering that the large magnetic moments of the Co adatoms are characterized by an out-of-plane easy axis, Kondo-screening is unlikely to occur~\cite{Otte:2008}, and enforces the view that the zero-bias anomalies result from spin-excitations. Additional simulations performed on Co adatoms on Cu and Ag(001) surfaces as well as Ti adatom on Ag(001), shown in Supplementary Figure~1, provide an additional evidence that spin-excitations are potentially present on other materials, giving rise to the experimentally observed zero-bias anomalies.

Grounding on a powerful theoretical framework based on relativistic time-dependent density functional and many-body perturbation theories, we obtain differential conductance spectra reproducing extremely well the measured data.
We systematically demonstrate the presence of spinarons, which are many-body bound-states emerging from the interaction of electrons and spin-excitations. While the self-energies quantifying the interaction of the electrons and spin-excitations are dynamical in nature and account for various correlation effects, it would be interesting to prospect in the future the impact of correlations (in the spirit of DFT + U~\cite{Anisimov:1997}) on the ground state properties, such as the magnetic anisotropy energy and subsequently on the excitation behavior of the investigated materials.
In general, our findings call for a profound change of our understanding of measured zero-bias anomalies of various nanostructures, which stimulates further theoretical developments permitting the ab-initio investigation of Kondo features, spinarons, spin-excitations and spin-orbit driven physics on equal footing.  

The one-to-one agreement between our first-principles spectra and the available  experimental ones strongly advocates for the importance of the spin-excitations in the interpretation of the origin of the zero-bias anomalies. X-ray magnetic circular dichroism (XMCD) experiments should help to confort our findings by unveiling the magnetic nature as well as the magnetic anisotropy energy of the investigated adatoms as done for Co adatoms on Pt(111)~\cite{gambardella:2003}. Surprisingly, this was, so far, not performed. Temperature-dependent and magnetic-field STM-based measurements were, to our knowledge, not reported, which is explained by the extreme difficulty to probe with enough resolution modifications induced in the rather broad spectral features. We conjecture that this might change in the near future, for example with electron-spin-resonance STM (ESR-STM)~\cite{Balatsky:2011,Baumann:2015} if realized on metallic substrates. In this work, various experimental setups were proposed, which would permit to further confirm our predictions. For instance, the theoretical spectra are spin-dependent and therefore the weight of each spin-channel to the total STM spectrum should depend on the spin-polarization of the tip. Furthermore, the application of a magnetic field is expected to increase the gap of the intrinsic spin-excitations, while a splitting is expected for Kondo features. However, the presence of the spinaron leads to an unconventional behavior, that is the excitation gap increases but the effective dip is not fixed and shifts to larger bias voltages. Currently, a few STM setup allow to reach large magnetic fields (e.g., 14 T and even 38 T), which would be enough to check our predictions. But even if those fields are not available, a reasonable alternative would be to use the proximity-induced effective magnetic field emerging from an adjacent magnetic adatom. Finally, one could tune down the magnetic anisotropy energy in order to reduce the amount of electron-hole excitations that are responsible for the broadening of the spin-excitations. This could be realized by attaching a non-magnetic atom such as Cu, for example, to Co adatom, after which the experimental investigation of the impact of temperature and magnetic fields would become more amenable. 

By opening a new perspective on low-energy spectroscopic features characterizing subnanoscale structures deposited on substrates, built upon the pioneering work of the STS community (see e.g. Refs.~\citenum{Madhavan:1998,Knorr:2002ig,Heinrich:2004,Ternes:2008}), our findings motivate new experiments exploring the interplay of temperature, proximity effects and response to an external magnetic field, which can help  identifying the real nature of the observed excitations and unravel the complexity and richness of the physics behind the spinaron.

\begin{methods}
Our first-principles approach is implemented in the framework of the scalar-relativistic full-electron Korringa-Kohn-Rostoker (KKR) Green function augmented self-consistently with spin-orbit interaction~\cite{Papanikolaou:2002,Bauer:2014}, where spin-excitations are described in a formalism based on time-dependent density functional theory (TD-DFT) ~\cite{Lounis:2010,Lounis:2011,Lounis:2014,Lounis:2015,Manuel:2015} including spin-orbit interaction. Many-body effects triggered by the presence of spin-excitations are approached via many-body perturbation theory~\cite{Schweflinghaus:2014,Schweflinghaus:2016,Ibanez:2017} extended to account for relativistic effects. The method is based on multiple-scattering theory allowing an embedding scheme, which is versatile for the treatment of nanostructures in real 
space. 
The full charge density is computed within the atomic-sphere approximation (ASA) and local spin density approximation (LSDA) is employed for the evaluation of the exchange-correlation potential~\cite{Vosko:1980}. We assume an angular momentum cutoff at $l_{\text{max}} = 3$ for the orbital expansion of the Green function and when extracting the local density of states a k-mesh of $300 \times 300$ is considered.   The Co adatoms sit on the fcc stacking site relaxed towards the surface by $20\%$ ($14\%$)  of the lattice parameter of the underlying  Au and Ag (Cu) substrates. 

After obtaining the ground-state electronic structure properties, the single-particle Green functions are then employed for the construction of the tensor of dynamical magnetic susceptibilities, $\underline{\chi}(\omega)$, within time-dependent density functional theory (TD-DFT)~\cite{Samir:2010,Samir:2015,Manuel:2015} including spin-orbit interaction. 
The susceptibility is obtained from a Dyson-like equation, which renormalizes the bare Kohn-Sham susceptibility, $\underline{\chi}_\text{KS}(\omega)$ as
\begin{equation}
\underline{\chi}(\omega) = \underline{\chi}_\text{KS}(\omega) 
+ \underline{\chi}_\text{KS}(\omega)\,\underline{\mathcal{K}}\,\underline{\chi}(\omega)\quad.
\label{TDDFT_RPA}
\end{equation}
$\underline{\chi}_\text{KS}(\omega)$ describes uncorrelated electron-hole excitations, while $\underline{\mathcal{K}}$ represents the exchange-correlation kernel, taken in adiabatic LSDA (such that this quantity is local in space and frequency-independent~\cite{Gross:1985}). 
The energy gap in the spin excitation spectrum is accurately evaluated using a magnetization sum rule~\cite{Samir:2010,Samir:2015,Manuel:2015}. 

\end{methods}

\begin{addendum}
\item[Data availability] All data needed to evaluate the conclusions in the paper are present in the paper and/or the Supplementary Information. Additional data related to this paper may be requested from the authors.

\item[Code availability] The KKR Green function code that supports the findings of this study is available from the corresponding author on reasonable request.

\item We thank  Markus Ternes, Alexander Weismann, Nicolas Lorente and Wolf-Dieter Schneider  
 for fruitful discussions. We are grateful to Michael Crommie, Lars Diekh\"oner, Peter Wahl, Alexander Schneider, Markus Ternes, Klaus Kern for sharing with us their original data measured with scanning tunneling microscopy. This work is supported by the European Research Council (ERC) under the European Union's Horizon 2020 research and innovation programme (ERC-consolidator grant 681405 — DYNASORE). We acknowledge the computing time granted by the JARA-HPC Vergabegremium and VSR commission on the supercomputer JURECA at Forschungszentrum Jülich.

\item[Authors contributions] S.L. initiated, designed and supervised the project. J.B. developed the theoretical ab-initio scheme accounting for spin-orbit interaction in the calculation of the many-body self-energies. J.B. performed the simulations and F. S. M. G. contributed to data post-processing. All authors discussed the results and helped writing the manuscript. 

\item[Competing Interests] The authors declare that they have no competing interests.

\end{addendum}

\bibliographystyle{naturemag}
\bibliography{mylib_paper.bib}
\newpage

\import{./}{supplemental.tex}

\end{document}

%% file: supplemental.tex
\makeatletter

\let\saved@includegraphics\includegraphics
\renewenvironment*{figure}{\@float{figure}}{\end@float}
\makeatother

\renewcommand{\figurename}{Supplementary Figure}
\renewcommand{\thefigure}{\textbf{\arabic{figure}}}

\renewcommand{\thetable}{\textbf{\arabic{table}}}

\renewcommand{\thesection}{Supplementary Note \arabic{section}} 
\renewcommand{\theequation}{S.\arabic{equation}}
\renewcommand{\refname}{Supplementary References}

\newcommand{\filipe}[1]{\textcolor{red}{\textbf{[Filipe: #1]}}}

\newcommand{\HH}{\mathcal{H}}
\newcommand{\RR}{\mathcal{R}}
\newcommand{\MM}{\mathcal{M}}
\newcommand{\AAA}{\mathcal{A}}
\newcommand{\NN}{\mathcal{N}}
\newcommand{\muB}{\mu_{\text{B}}}
\newcommand{\upup}{\uparrow\uparrow}
\newcommand{\eqand}{\quad \textnormal{and} \quad}
\newcommand{\eqdot}{\quad .}
\newcommand{\eqcomma}{\quad ,}
\newcommand{\onehalf}{\frac{1}{2} }

\bibliographystyle{naturemag}


\begin{center}
     \Large{\textbf{Supplementary Information: A new view on the origin of zero-bias anomalies of Co atoms atop noble metal surfaces}}
    \vspace{1cm}
    
    \large{Juba Bouaziz, Filipe S. M. Guimar\~aes and Samir Lounis}
    \vspace{0.5cm}
    
    \large{\textit{Peter Gr\"unberg Institut and Institute for Advanced Simulation, 
Forschungszentrum  J\"ulich and JARA, 52425 J\"ulich, Germany}}
\end{center}

\newpage 

\begin{figure}
    \centering
    \includegraphics[width=0.7\columnwidth]{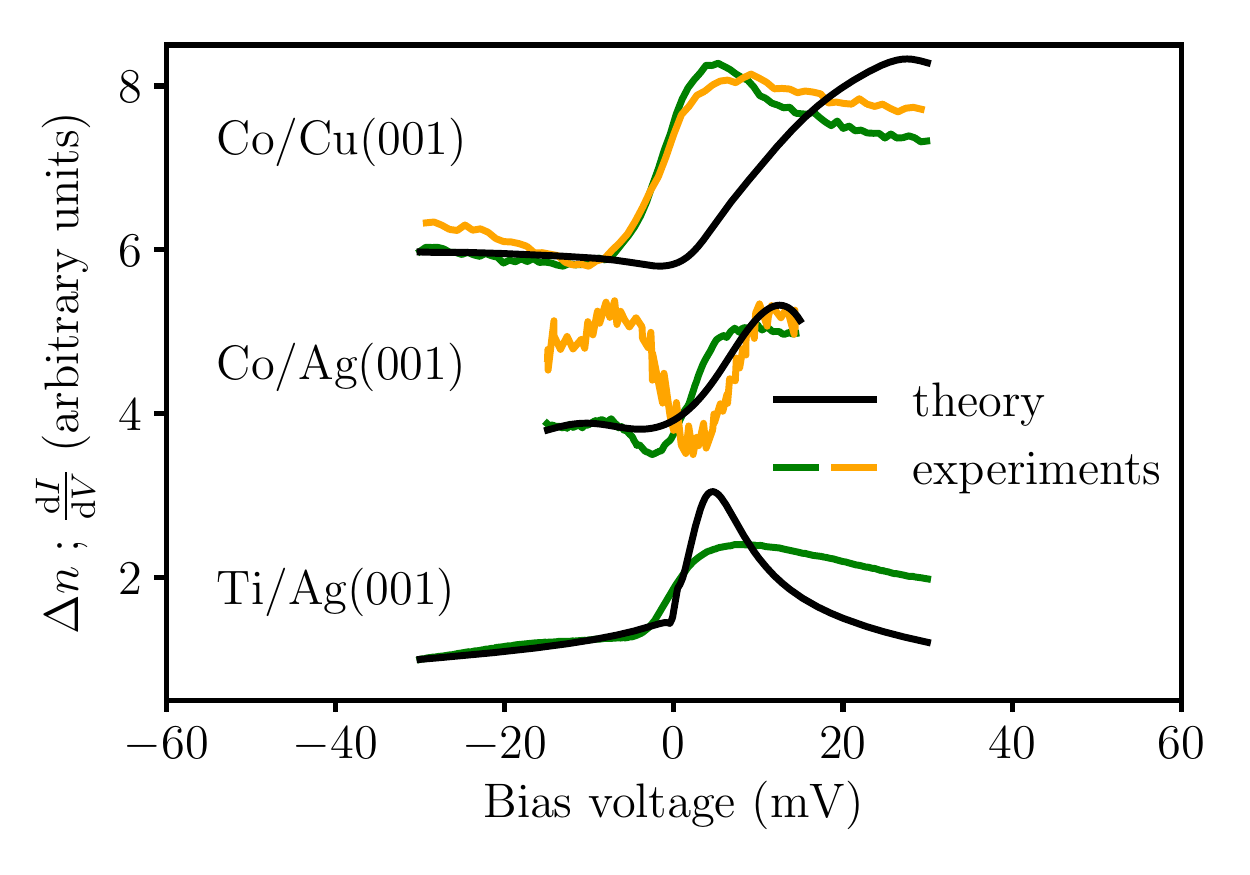}
    \caption{\textbf{Zero-bias anomalies calculated and measured on Co adatoms on Cu and Ag(001) surfaces as well as Ti adatom on Ag(001) surface.} \small{For Co/Cu(001), the agreement with the experimental data~\cite{Knorr:2002ig,Wahl:2007kf} is not as perfect as it is on the (111) surfaces. We recover, however, the observed step-like behavior. A reason could be the underestimation of the magnetic anisotropy energy of Co on Cu(001), which can shift slightly the spectrum around. For Co/Ag(001), the agreement is rather good when compared to the data of Ternes et al.~\cite{Ternes:2008}. It is interesting to notice, however, that the measurements of Wahl et al.~\cite{Wahl:2004jy} lead to different spectra. The disagreement between the two experimental data can be surprising at first sight. We conjecture that this can be induced by the presence of hydrogen or by the difference in the probing tip, which can change the shape of the features. This strongly motivates further experimental investigations. The case of of Ti/Ag(001) measured by measured by Nagaoka et al.~\cite{Nagaoka:2002} is rather reasonably described by our theory. Our zero-bias anomaly is sharper than the experimental one. Overall, the agreement between theory and experiment is rather good on the (001) surfaces, which indicates that spin-excitations, as discussed in the main text, is a plausible origin of the low-energy features on the (001) surfaces of Cu and Ag, similarly to the (111) surfaces. Note that the position of the  theoretically obtained features hinges on the ability to  evaluate the magnetic anisotropy energy of the adatoms, which is not always trivial. The experimental data  adapted with permission from Refs.~\citenum{Knorr:2002ig,Nagaoka:2002,Wahl:2004jy,Wahl:2007kf,Ternes:2008}. Copyright (2002), (2004) and (2007) by the American Physical Society and (2008) by IOP Publishing.}}
    \label{fig:001}
\end{figure}

\begin{figure}
    \centering
    \includegraphics[width=0.6\columnwidth]{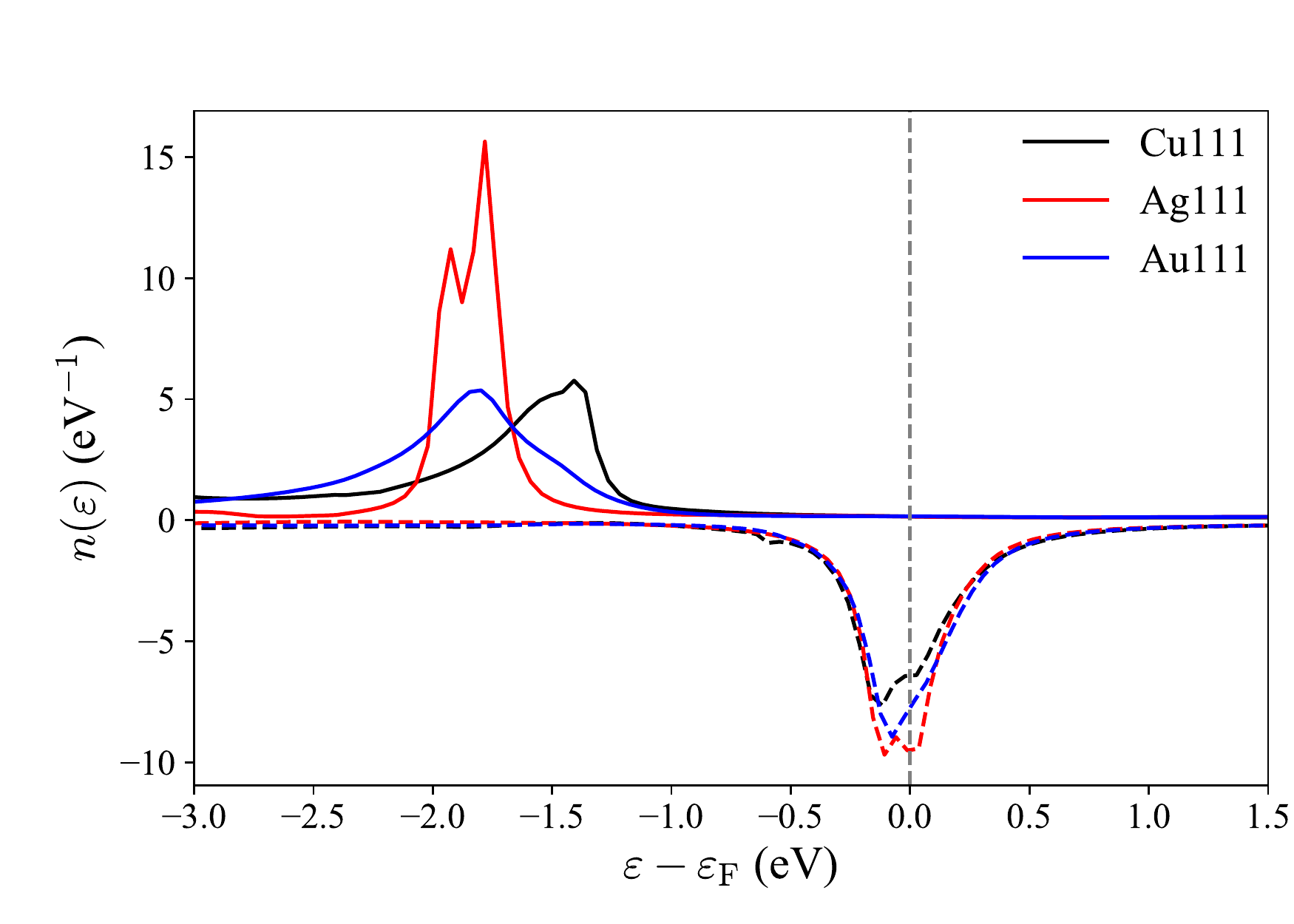}
    \caption{\textbf{Total spin-polarized local density of states of a Co adatom on Cu, Ag and Au(111) surfaces.} The minority-spin state (lower panel) for Co on Ag(111) is sharper due to the weaker strength of the hybridization of the electronic states of the adatom and the substrate, when compared to Au or Cu(111).}
    \label{fig:Co_LDOS}
\end{figure}

\begin{figure}
    \centering
    \includegraphics[width=\textwidth]{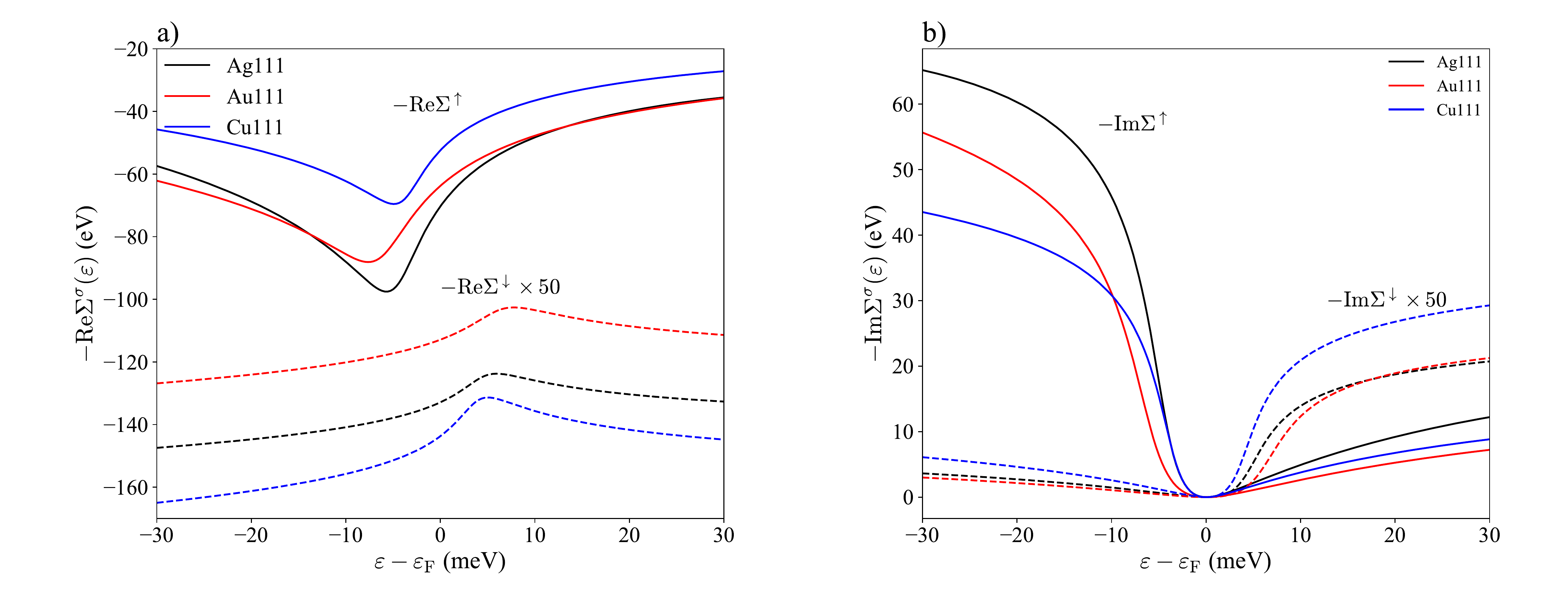}
    \caption{\textbf{Comparison of the spin-resolved imaginary and real parts of the trace of the Co adatoms self-energies on the three surfaces Cu, Ag, Au(111).} Similarly to Au(111) surface (shown also in the main Figure 2b), the steps characterizing the imaginary part of the self-energies are asymmetric. The one generated at negative bias voltage corresponding to the majority-spin channel is the largest on Ag. This is induced by the large minority-spin local density of states (see Supplemental Figure~\ref{fig:Co_LDOS}), which defines the height of the step.   }
    \label{fig:self-energies}
\end{figure}

\begin{figure}[ht!]
    \centering
    \includegraphics[width=\textwidth]{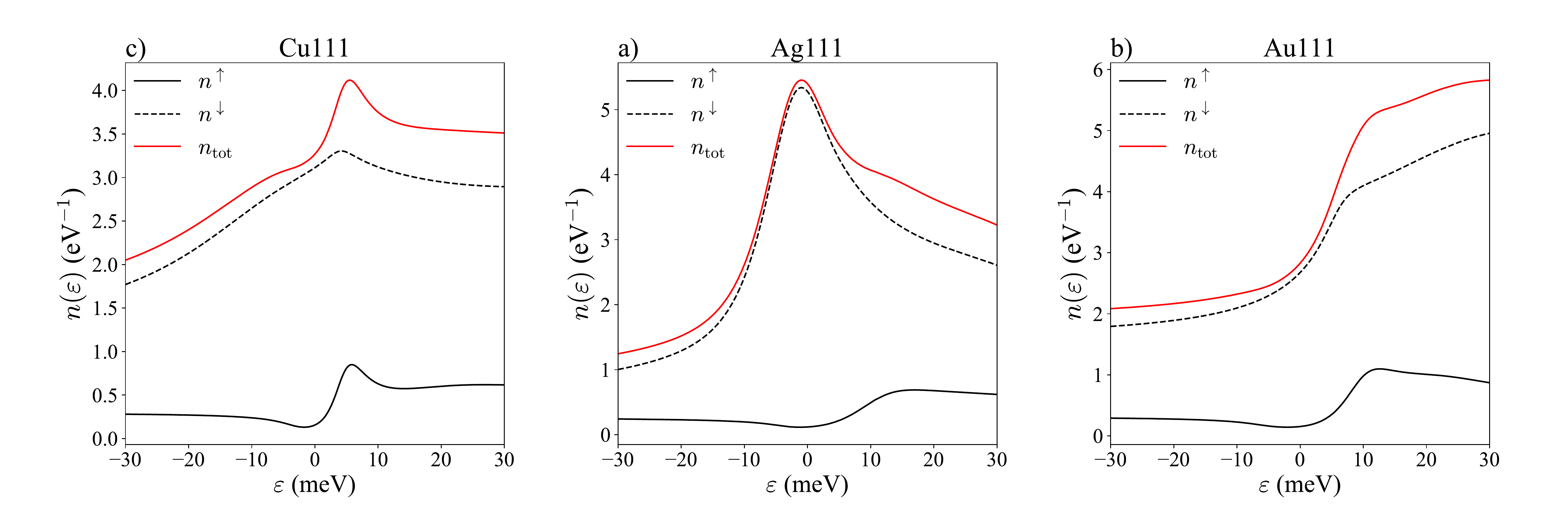}
    \caption{\textbf{Comparison of the renormalized local density of states of Co adatoms on the three surfaces Cu, Ag, Au(111).} Black full (dashed) lines correspond to the majority-spin (minority-spin states), while the full red lines depict the total density of states. The theoretical inelastic spectra are calculated in the vacuum as a result of the decay of  adatoms electronic states. Therefore the shape of the signature of the spin-excitations and of the spinaron do not have to be the same in the adatom and in the vacuum.    }
    \label{fig:renormalized_ldos}
\end{figure}

\begin{figure}[ht!]
    \centering
    \includegraphics[width=\textwidth]{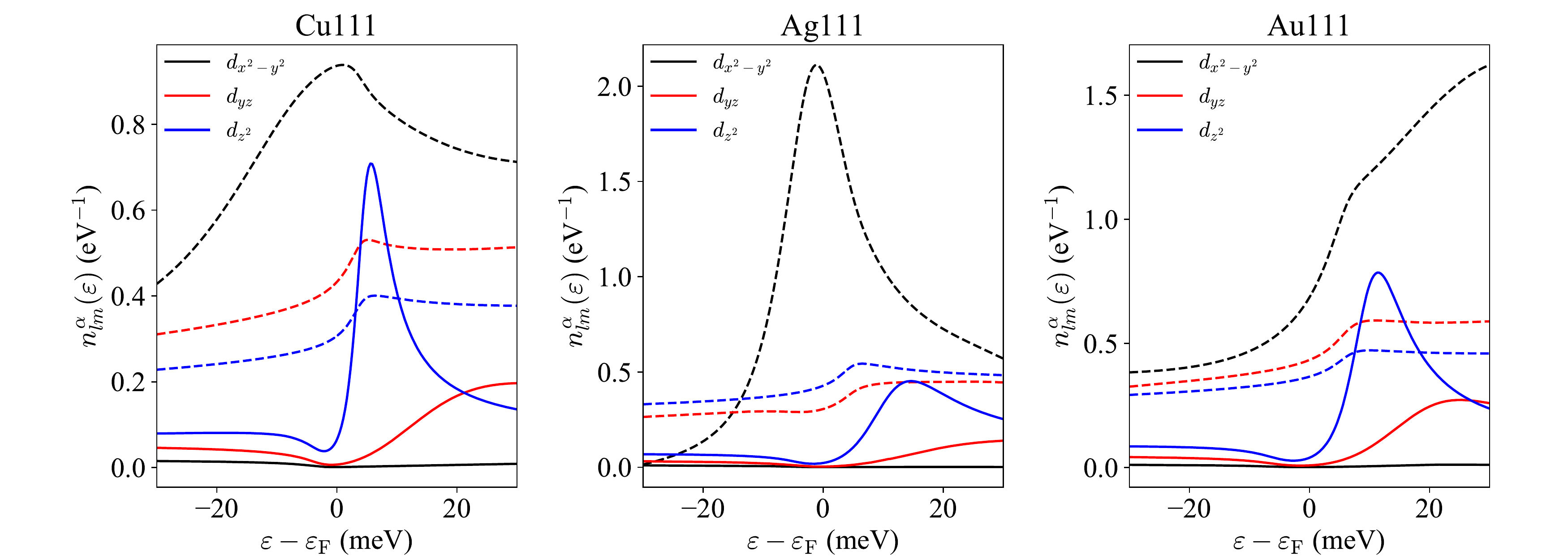}
    \caption{\textbf{Orbital-resolved renormalized local density of states of Co adatoms on the three surfaces Cu, Ag, Au(111).} Full (dashed) lines correspond to the majority-spin (minority-spin) channel. The spinaron emerges from the majority-spin $d_{z^2}$ orbital (full blue line).}
    \label{fig:reno_lm_ldos}
\end{figure}

\begin{figure}[ht!]
    \centering
    \includegraphics[width=0.6\textwidth]{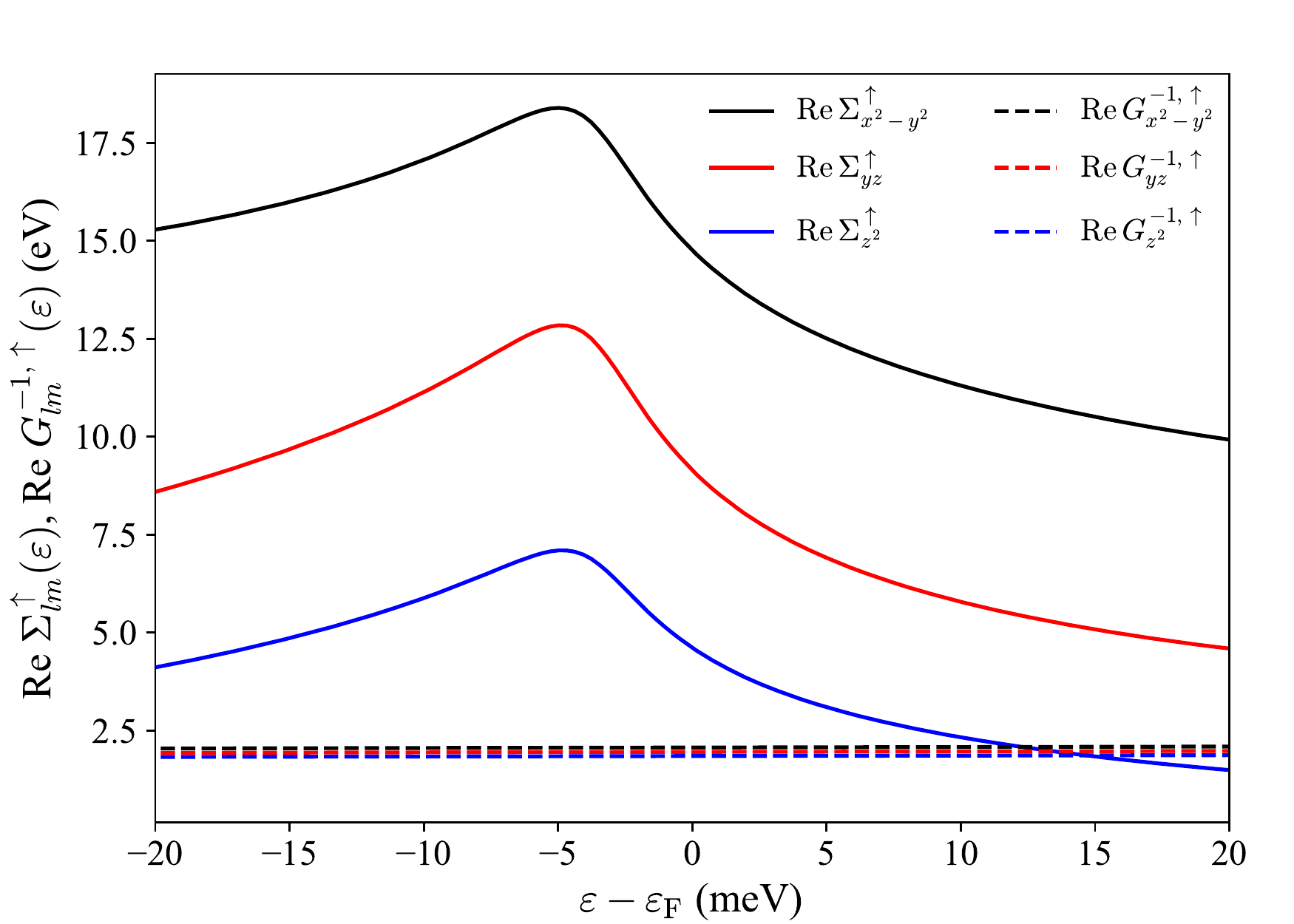}
    \caption{\textbf{Orbital-resolved real part of inverse Green function and self-energy of the majority spin channel of Co adatoms on Cu(111) surface.} The intersection occurring in the $d_{z^2}$-orbital leads to a vanishing denominator of the Eq~1 of the main text and, therefore, to the spinaron. The contribution of $\text{Im}[\underline{G}(\varepsilon)]\text{Im}[\underline{\Sigma}(\varepsilon)]$ (and other interference effects) shifts the spinaron to lower energies.}
    \label{fig:re_self_lm_cu111}
\end{figure}

\begin{figure}
	\centering
	\includegraphics[width=0.8\textwidth]{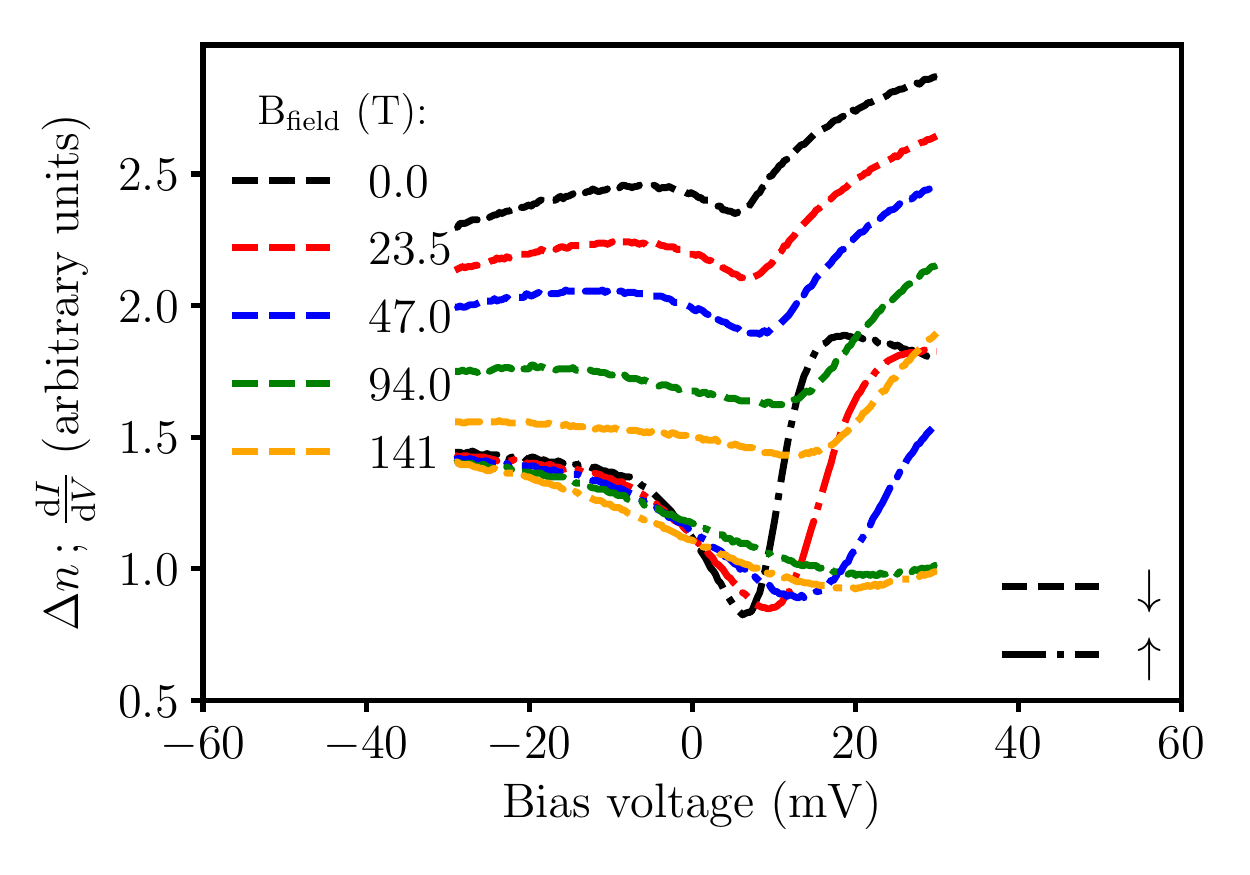}
	\caption{\textbf{Theoretical scanning tunneling spectroscopy spectra probing the differential conductance of a Co adatom on Au(111).} After applying a magnetic field, the spin-resolved spectra of the zero-bias signals (shown in Fig.~3C of the main text) experience various changes. The signature of the spin-excitation in the minority-spin channel moves to larger energies, as expected. 
	The spinaron, living in the majority-spin channel, also shifts to higher energies. 
	This occurs as a result of the increase of the spin-excitation gap upon application of the magnetic field. 
	The intrinsic spin-excitation expected at negative bias voltage should shift to lower energies as well.
	However, it is difficult to track its low signal, that is caused by the large imaginary part of the corresponding self-energy, which considerably lowers the lifetime of the majority-spin electrons. 
	In other words, the expected feature is much broader than the one occurring in the minority-spin channel. 
	Overall, we notice that all features experience a decrease in their lifetimes since more electron-hole excitations are allowed when increasing the excitation energy.
	Summing up both spin-channels, as done in Fig.~3C, leads to two steps with a dip in-between. The dip shifts to higher energies, while the gap increases in magnitude after application of the magnetic field.
} \label{sp_field}
\end{figure}

\clearpage

\section{Description of the formalism used to describe the self-energy}

\subsection{General form of the self-energy.}
\label{SE_freq_str} Utilizing many-body perturbation theory, the self-energy describing the interaction of electrons and spin-excitations can be written in the following form:
\begin{equation}
\begin{split}
\Sigma^{\sigma\sigma'}(\mathbf{r},\mathbf{r}^{\prime},\varepsilon)  = -\frac{U(\mathbf{r})\,U(\mathbf{r}^{\prime})}{\pi} \sum_{ss^\prime} \bigg\{&\int_{0}^{\infty}d\omega\,\text{Im}\left[G^{ss'}(\mathbf{r},\mathbf{r}^{\prime},\varepsilon+\omega)\,\chi^{\sigma s,s'\sigma'}(\mathbf{r},\mathbf{r}^{\prime},\omega)\right] \\
-&\int_{0}^{\varepsilon_\text{F}-\varepsilon} d\omega\,\text{Im}\left[G^{s s'}(\mathbf{r},\mathbf{r}^{\,\prime},\varepsilon+\omega)\right]\left[\chi^{\sigma s,s'\sigma'}(\mathbf{r}^{\prime},\mathbf{r},\omega)\right]^{*}
\bigg\}\quad,
\end{split}
\label{SE_general}
\end{equation}
which is a generalized form~\cite{Fransson:2015} of the self-energy described in Ref.~\citenum{Schweflinghaus:2014}, where spin-orbit interaction is included, leading to possible contributions (in spin-space) of off-diagonal elements of the Green functions and of the tensor of the  dynamical spin-susceptibility. 
$\chi^{\sigma s,s^{\prime}\sigma^{\prime}}(\omega)$ is the Fourier transform of the spin-susceptibility $\chi^{\sigma s,s^{\prime}\sigma^{\prime}}(t) = -\iu\,\Theta(t)\langle [\hat{S}_{\sigma s}(t),\hat{S}_{s^{\prime}\sigma^{\prime}}]\rangle$, and $\sigma,\sigma',s,s'=\{\uparrow,\downarrow\}$. The latter can be identified as the susceptibility obtained from time-dependent density functional theory (TD-DFT) while  $U(\vec{r})$ represents the exchange-correlation kernel, evaluated in the adiabatic local spin-density approximation (ALSDA)~\cite{Schweflinghaus:2014}. 
The self-energy consists then on two kinds of elements: spin-diagonal and off-diagonal contributions.

\subsection{Diagonal spin-components of the self-energy and  approximations.} The diagonal spin-terms of the self-energy are $\Sigma^{\uparrow\uparrow}$ and  $\Sigma^{\downarrow\downarrow}$.
The former is schematically given by
\begin{equation}
\begin{split}
\Sigma^{\uparrow\uparrow}(\varepsilon) \propto& \sum_{ss^{\prime}} \chi^{\uparrow s,s^{\prime}\uparrow}(\omega)\,G^{ss^{\prime}}(\omega+\varepsilon)\\
\propto&\chi^{\uparrow \uparrow,\uparrow\uparrow}(\omega)\,G^{\uparrow\uparrow}(\omega+\varepsilon)
+\chi^{\uparrow \uparrow,\downarrow\uparrow}(\omega)\,G^{\uparrow\downarrow}(\omega+\varepsilon)\\
&+\chi^{\uparrow \downarrow,\uparrow\uparrow}(\omega)\,G^{\downarrow\uparrow}(\omega+\varepsilon)
+\chi^{\uparrow \downarrow,\downarrow\uparrow}(\omega)\,G^{\downarrow\downarrow}(\omega+\varepsilon)
\end{split}
\end{equation}
The bias voltages that are addressed experimentally to probe spin-excitations of adatoms via inelastic scanning tunneling spectroscopy (ISTS) are located in the \SI{}{\milli\electronvolt} range, where the transverse spin-excitations are located.
These modes are encrypted in the transverse (spin-flip) block of the susceptibility.
For the systems we investigate, the off-diagonal elements of the susceptibility tensor, i.e. the  transverse-longitudinal block, is found negligible, i.e. $\chi^{\uparrow \uparrow,\downarrow\uparrow}(\omega)\simeq\chi^{\uparrow \downarrow,\uparrow\uparrow}(\omega)\simeq0$.
The longitudinal component $\chi^{\uparrow \uparrow,\uparrow\uparrow}(\omega)$ encodes excitations located at higher frequencies (\SI{}{\electronvolt} range), so its contribution is neglected for the investigated frequencies. Therefore, $\Sigma^{\uparrow\uparrow}(\varepsilon)$ is simply given by 
\begin{equation}\label{se_spin_up}
\begin{split}
\Sigma^{\uparrow\uparrow}(\varepsilon) &\propto \chi^{+,-}(\omega)\,G^{\downarrow\downarrow}(\omega+\varepsilon)\quad,
\end{split}
\end{equation}
and similarly
\begin{equation}\label{se_spin_down}
\begin{split}
\Sigma^{\downarrow\downarrow}(\varepsilon) &\propto \chi^{-,+}(\omega)\,G^{\uparrow\uparrow}(\omega+\varepsilon)\quad,
\end{split}
\end{equation}
where the usual short notation $\chi^{+,-}(\omega)=\chi^{\uparrow \downarrow,\downarrow\uparrow}(\omega)$ and $\chi^{-,+}(\omega)=\chi^{\downarrow \uparrow,\uparrow\downarrow}(\omega)$ has been employed.
We note that our extended approach includes the intrinsic gap opening in the spin-excitation spectrum due to the breaking of rotational symmetry induced by spin-orbit interaction. 

\subsection{Off-diagonal spin-components of the self-energy and approximations.} 
The spin off-diagonal contribution to the self-energy  reads
\begin{equation}
\Sigma^{\downarrow\uparrow}(\varepsilon) \propto 
 \chi^{\downarrow \uparrow,\uparrow\uparrow}(\omega)\,G^{\uparrow\uparrow}(\omega+\varepsilon)
+\chi^{\downarrow \uparrow,\downarrow\uparrow}(\omega)\,G^{\uparrow\downarrow}(\omega+\varepsilon)
+\chi^{\downarrow \downarrow,\uparrow\uparrow}(\omega)\,G^{\downarrow\uparrow}(\omega+\varepsilon)
+\chi^{\downarrow \downarrow,\downarrow\uparrow}(\omega)\,G^{\downarrow\downarrow}(\omega+\varepsilon)\ .
\end{equation}
As aforementioned, the off-diagonal longitudinal-transverse block and the longitudinal contributions of the susceptibility tensor have negligible contributions for the investigated adatoms.
The remaining term involving  the component $\chi^{-,-}(\omega)=\chi^{\downarrow \uparrow,\downarrow\uparrow}(\omega)$ vanishes in systems with $C_{3v}$ symmetry or higher, as the ones investigated in our work~\cite{Guimaraes:2017cv}.
Therefore, the off-diagonal components of the self-energy $\Sigma^{\downarrow\uparrow}(\varepsilon)$ and $\Sigma^{\uparrow\downarrow}(\varepsilon)$ can be safely neglected for the specific cases investigated in the main manuscript. 

\subsection{Forms of the self-energies shown in the manuscript.} 
The self-energy used in practical calculations has a spin-diagonal form. 
For an applied bias voltage $V$, the real part of the spin-diagonal components of the self-energy read 
\begin{equation}
\begin{split}
\text{Re}\,\Sigma^{\sigma\sigma}(\mathbf{r},\mathbf{r}^{\prime},\varepsilon_\text{F}+V)  =& -\frac{U(\mathbf{r})\,U(\mathbf{r}^{\prime})}{\pi}\bigg\{\int_{0}^{\infty}d\omega\,\text{Im}\left[G^{\bar{\sigma}\bar{\sigma}}(\mathbf{r},\mathbf{r}^{\prime},\varepsilon_\text{F}+V+\omega)\,\chi^{\sigma\bar{\sigma},\bar{\sigma}\sigma}(\mathbf{r},\mathbf{r}^{\prime},\omega)\right] \\ 
& - \int_{0}^{-V} d\omega\,\text{Im}\left[G^{\bar{\sigma}\bar{\sigma}}(\mathbf{r},\mathbf{r}^{\prime},\varepsilon_\text{F}+V+\omega)\right]\text{Re}\left[\chi^{\sigma\bar{\sigma},\bar{\sigma}\sigma}(\mathbf{r}^{\prime},\mathbf{r},\omega)\right]\bigg\}\quad.
\end{split}
\label{SE_real}
\end{equation}
Note that the first term in Eq.~\ref{SE_real} involves a frequency integration up to infinity. 
In practice, the evaluation of the integral is done in two steps: First, the magnetic susceptibility is computed up to a cut-off frequency $\omega_\text{cut}= \SI{100}{\milli\electronvolt}$. The value of the cut-off is chosen after systematic convergence calculations.
Second, for the remaining range, we make use of the high-frequency expansion of the spin-flip susceptibility~\cite{Giuliani:2005cb}
\begin{equation}
\begin{split}
\lim_{\omega\rightarrow+\infty}\text{Re}\,\chi_{+-}(\omega)&\approx \frac{C^\text{Re}}{\omega} + ...\quad,\\
\lim_{\omega\rightarrow+\infty}\text{Im}\,\chi_{+-}(\omega)&\approx  \left(\frac{C^\text{Im}}{\omega^2}\right) + ...\quad,
\end{split}
\end{equation}
where $C^\text{Re} = \text{Re}\,\chi_{+-}(\omega_\text{cut})\,\omega_\text{cut}$ and $C^\text{Im} = \text{Im}\,\chi_{+-}(\omega_\text{cut})\,\omega^2_\text{cut}$.

The imaginary part of the self-energy is simply given by
\begin{equation}
\begin{split}
\text{Im}\,\Sigma^{\sigma\sigma}(\mathbf{r},\mathbf{r}^{\prime},\varepsilon_\text{F}+V) & = -\frac{U(\mathbf{r})\,U(\mathbf{r}^{\prime})}{\pi} \int_{0}^{-V} d\omega\,\text{Im}\left[G^{\bar{\sigma}\bar{\sigma}}(\mathbf{r},\mathbf{r}^{\prime},\varepsilon_\text{F}+V+\omega)\right]\text{Im}\left[\chi^{\sigma\bar{\sigma},\bar{\sigma}\sigma}(\mathbf{r}^{\prime},\mathbf{r},\omega)\right].
\end{split}
\label{SE_imag}
\end{equation}
For the bias voltages applied experimentally ($\sim \SI{}{\milli\electronvolt}$), the imaginary part of the Green function of the investigated adatoms depends weakly on the energy.
As a result, the imaginary part of the self-energy presents a step as function of the bias voltage when integrating over the peak position of  the susceptibility. Thus the position of the step is the hallmark of the excitation energy.

\subsection{Projection basis for the  Green function and susceptibility}
The aforementioned self-energy  is given as 
a convolution of the single particle Green function and the spin-spin susceptibility. 
In practice, these quantities are evaluated from the Korringa-Kohn-Rostoker (KKR) Green function approach using the atomic sphere approximation (ASA)~\cite{Papanikolaou:2002,Bauer:2014,Lounis:2011}. 
Within this method, the space is partitioned into cells centred around atomic sites: 
\begin{equation}
{G}^{\sigma\sigma^{\prime}}_{i,j}(\mathbf{r},\mathbf{r}^{\prime},\varepsilon) = \sum_{L,L^\prime}Y_{L}(\hat{\mathbf{r}})\,{G}^{\sigma\sigma^{\prime}}_{iL,jL^{\prime}}
({r},{r}^{\prime},\varepsilon)
\,Y_{L^\prime}(\hat{\mathbf{r}}^{\prime})\quad,
\label{kkr_gf}
\end{equation}
the distance $r$ $({r}^{\,\prime})$ is measured from the atomic site $i$ $(j)$. 
The angular dependence is expanded in real spherical harmonics $Y_{L}(\hat{\mathbf{r}})$, 
with $L=\{l,m\}$ being the azimuthal and magnetic quantum numbers, respectively. 
Furthermore, to access complex quantities such as response functions a projection 
basis is introduced in practice~\cite{Lounis:2011,Manuel:2015}. 
The latter is built from the spin-independent regular solutions ${R}_{il}(r,\varepsilon)$:
\begin{equation}
\phi_{ilb}(r) = \frac{{R}_{il}(r,\varepsilon_\text{b})}{\sqrt{\int_{0}^{R} 
dr\,r^{2}\,{R}_{il}(r,\varepsilon_\text{b})^{2}}}\quad. 
\label{projection_basis}
\end{equation}
$\varepsilon_\text{b}$ represents a set of energies chosen in the the valence states energy range. 
The radial part of the Green function is the expressed in the the projection basis as 
\begin{equation}\label{Proj_Gf}
{G}^{\sigma\sigma^{\prime}}_{iL,jL^{\prime}}
({r},{r}^{\prime},\varepsilon) = \sum_{bb^{\prime}}\,\phi_{ilb}(r)\,{G}^{\sigma\sigma^{\prime}}_{iLb,jL^{\prime}b^{\prime}}
(\varepsilon)
\,\phi^{}_{jl^{\prime}b^{\prime}}(r^{\prime})\quad.
\end{equation} 
$b$ and $b^{\prime}$ define the basis indices. 

The convolution of two single-particle Green functions gives rise to the Kohn-Sham magnetic 
susceptibility~\cite{Lounis:2011,Manuel:2015}.    
Once more, to reduce the computational costs, we introduce a mixed product basis 
$\Phi_{iLb}(x)$ by contracting the product of the Green function basis~\cite{Aryasetiawan:1994,Manuel:2015}:
\begin{equation}
\sum_{b}C^{iLb}_{L_1b_1L_2b_2}\Phi_{iLb}(r) = C^{L}_{L_1L_2}\phi_{il_{1}b_{1}}(r)\,\phi_{il_{2}b_{2}}(r)\quad.
\end{equation}
$C^{L}_{L_1L_2}$ and $C^{iLb}_{L_1b_1L_2b_2}$ represent gaunt and generalized Gaunt 
coefficients, respectively. In this new basis, the response function reads:
\begin{equation}\label{chi_expansion}
\chi^{\alpha\beta}_{ij}(\mathbf{r},\mathbf{r}^{\prime},\omega) = \sum_{Lb,L^\prime b^\prime} 
\Phi_{iLb}(r)\,Y_{L}(\hat{\mathbf{r}})\,\chi^{\alpha\beta}_{iLb,jL^\prime b^\prime}(\omega)\,
\Phi_{jL^\prime b^\prime}(r^\prime)\,Y_{L}(\hat{\mathbf{r}}^\prime)\quad. 
\end{equation}

Finally, we provide the basis representation of the self-energy, which is  schematically proportional to:
$
\Sigma_{ij}(\mathbf{r},\mathbf{r}^{\,\prime},\varepsilon) \propto U_{i}(\mathbf{r})\,G_{ij}(\mathbf{r},\mathbf{r}^{\prime},\varepsilon)\,
\chi_{ij}(\mathbf{r},\mathbf{r}^{\prime},\varepsilon^{\prime})\,U_{j}(\mathbf{r}^{\prime})$.
Following the expansion chosen for the Green function as defined in Eq.~\ref{projection_basis}, the self-energy reads:
\begin{equation}
{\Sigma}_{i,j}(\mathbf{r},\mathbf{r}^{\prime},\varepsilon) = \sum_{Lb,L^\prime b^{\prime}}Y_{L}(\hat{\mathbf{r}})\,\phi_{ilb}(r)\,{\Sigma}_{iLb,jL^{\prime}b^{\prime}}(\varepsilon)
\,\phi_{jl^{\prime}b^{\prime}}(r^{\prime})\,Y_{L^\prime}(\hat{\mathbf{r}}^{\,\prime})\quad,
\end{equation} 
with the self-energy components given by
\begin{equation}
\begin{split}
{\Sigma}_{iL_1b_1,jL_2b_2}(\varepsilon) = \sum_{Lb,L^\prime b^{\prime}}
\sum_{\substack{L_3b_3,L_4b_4\\L_5b_5,L_6b_6}}C^{iL_3b_3}_{L_1b_1Lb}
C^{jL_4b_4}_{L_2b_2L^\prime b^\prime} U_{i,L_3b_3,L_5b_5}\, {G}^{}_{iLb,jL^{\prime}b^{\prime}}(\varepsilon)
\,\chi_{iL_5b_5,jL_6b_6}(\varepsilon^{\prime})\,U_{j,L_6b_6,L_4b_4}
\end{split}
\end{equation}
Finally, assuming that the kernel is diagonal in the basis, and using 
the spherical average of the magnetic susceptibility and of the exchange-correlation kernel~\cite{Manuel:2015}, 
the previous equation reduces to
\begin{equation}
{\Sigma}_{iL_1b_i,jL_2b_j}(\varepsilon) 
 = \sum_{\substack{bb^{\prime}b_3b_4}}C^{i0b_3}_{L_1b_iL_1b1}
\,C^{j0b_4}_{L_2b_jL_2 b_2}\,U_{i,0b_3}\,{G}_{iL_1b_1,jL_2b_2}(\varepsilon)
\,\chi_{i0b_3,j0b_4}(\varepsilon^{\prime})\,U_{j,0b_4}\quad.
\end{equation}

\section{Simplified scheme with an auxiliary magnetic field instead of magnetic anisotropy}

We note that the case of Co adatom on Cu(111) was addressed in the preliminary work reported in Ref.~\citenum{Schweflinghaus:2014}, where the self-energies of the adatoms were computed in a very simplified approach compared to the current work. The results are, however, in line with those obtained with the more accurate formalism described in our manuscript. In Ref.~\citenum{Schweflinghaus:2014}, a rudimentary projection basis (based on regular scattering solutions that are computed at the Fermi energy) was utilized, which is not that precise to describe various spin-dynamics properties. More importantly, it was lacking the impact of the spin-orbit coupling. The simplest implication of this interaction is to open a gap of a few \SI{}{\milli\electronvolt} in the susceptibility, which has to be precisely described.
Instead of spin-orbit coupling, a magnetic field was utilized to effectively open a gap in the spin-excitations spectra. 
The former has profoundly different implications on the dynamical susceptibilities compared to those induced by the latter. 
The main ingredient is the definition of the correct ground state of the magnetic orientation with respect to the lattice, which is not known when magnetic fields are used. 
Additionally, the magnitude of the field is arbitrary and has a different physical origin. 
These aspects induce drastic modifications in the properties of the susceptibilities --- and therefore in the self-energy and the resulting renormalized electronic structure: the shape and weight of the spin-flip relativistic responses, $\chi^{+-}$ and $\chi^{-+}$, at positive and negative frequencies (both imaginary and real parts) change dramatically if the moment is lying in an easy-plane or along an easy axis, depending also if it is in the ground state or in a metastable state.
Consequently, it is impossible to make any reasonable claim and systematic interpretation of the experimental data of Co on the three noble metallic substrates (Cu, Ag, Au) from such a poor's man approach.
These issues were solved by our current approach.


